\documentclass{appolb}
\usepackage{graphicx}

\usepackage{dcolumn}
\usepackage{bm}
\usepackage{slashed}
\usepackage{amsmath,graphicx}
\usepackage[colorlinks=true,linktocpage=true,linkcolor=blue,citecolor=blue]{hyperref}
\usepackage{float}
\usepackage{nicefrac}
\usepackage[normalem]{ulem}
\usepackage{amsmath}
\usepackage{subfigure}

\newcommand{\beq}{\begin{eqnarray}}
\newcommand{\eeq}{\end{eqnarray}}

\newcommand{\bel}[1]{\begin{eqnarray}\label{#1}}
\newcommand{\eel}{\end{eqnarray}}

\newcommand{\rf}[1]{Eq.~(\ref{#1})}

\newcommand{\rfn}[1]{(\ref{#1})}

\newcommand{\p}{\partial}

\newcommand{\f}[2]{\frac{#1}{#2}}
\newcommand{\onehalf}{{\nicefrac{1}{2}}}

\newcommand{\ed}{{\varepsilon}}       

\def\gmunu{g^{\mu\nu}}

\def\TmnU{T^{\mu \nu}}

\def\n0{n_{(0)}}
\def\e0{\varepsilon_{(0)}}
\def\P0{P_{(0)}}
\def\s0{s_{(0)}}

\def\fplusrsxp{f^+_{rs}(x,p)}

\def\umU{u^\mu}

\def\unu{u^\nu}

\def\pmu{p^\mu}

\def\omnL{\omega_{\mu\nu}}
\def\omnU{\omega^{\mu\nu}}

\def\ubarrp{{\bar u}_r(p)}

\def\usp{u_s(p)}

\def\vbarsp{{\bar v}_s(p)}

\def\vrp{v_r(p)}

\def\g5{\gamma_5}

\def\fminusrsxp{f^-_{rs}(x,p)}

\begin{document}
\title{Perfect-fluid hydrodynamics with constant acceleration along the stream lines and spin polarization%
}
\author{Wojciech Florkowski
\address{Institute of Nuclear Physics, PL-31-342 Krakow, Poland \\
Jan Kochanowski University, PL-25-406 Kielce, Poland}
\\ \medskip
{Enrico Speranza
}
\address{Institute for Theoretical Physics, Goethe University, \\
D-60-438 Frankfurt am Main, Germany}
\\ \medskip
{Francesco Becattini
}
\address{Universit`a  di  Firenze  and  INFN  Sezione  di  Firenze,  Florence,  Italy}
}

\maketitle
\begin{abstract}
A global equilibrium state of a spin polarized fluid that undergoes constant acceleration along the stream lines 
is described as a solution of recently introduced perfect-fluid hydrodynamic equations with spin $\onehalf$. 
\end{abstract}

\section{Introduction}

Recent measurements of the spin polarization of $\Lambda$ hyperons in heavy-ion collisions \cite{Abelev:2007zk, STAR:2017ckg} triggered broad interest in the relation between polarization and fluid vorticity.  At local thermodynamic equilibrium, the spin polarization tensor is directly related to so called thermal vorticity~\cite{Becattini:2013fla}, provided that local equilibrium is defined without any constraint on the spin tensor~\cite{preparation}. The latter is defined by the expression $\varpi_{\mu\nu} = -\f{1}{2} (\p_\mu \beta_\nu - \p_\nu \beta_\mu)$, where $\beta^\mu = u^\mu/T$, with $T$ and $u^\mu$ being the system's local temperature and four-velocity, respectively. The other recent topics related to polarization include the kinetics of spin~\cite{Gao:2012ix, Fang:2016vpj, Fang:2016uds}, anomalous hydrodynamics~\cite{Son:2009tf, Kharzeev:2010gr,Sadofyev:2010pr,Neiman:2010zi}, and the Lagrangian formulation of hydrodynamics~\cite{Montenegro:2017rbu, Montenegro:2017lvf}. Several issues connected with the global and local spin polarization have been recently reviewed in~\cite{Wang:2017jpl}. 

While the relation between polarization and thermal vorticity is firmly established in global equilibrium situations, it is still under investigations if finite polarization may exist in a properly defined local thermodynamic equilibrium situation without thermal vorticity. In other words, it is possible that in the most general relativistic fluid at local thermodynamic equilibrium the spin polarisation tensor $\omega_{\mu\nu}$, which is proportional at the leading order to $\varpi_{\mu\nu}$, does not coincide with $-\f{1}{2} (\p_\mu \beta_\nu - \p_\nu \beta_\mu)$. Steps in this direction have been taken in Refs.~\cite{Florkowski:2017ruc,Florkowski:2017dyn},  where the framework of perfect-fluid hydrodynamics with spin $\onehalf$ was formulated. It  was demonstrated that global equilibrium states with spin polarization and vorticity may be interpreted as stationary solutions of the hydrodynamic equations with spin introduced in Ref.~\cite{Florkowski:2017ruc}.

Interestingly, global equilibrium states not only include the cases with rotation but also with constant acceleration along the fluid stream lines. This case has been recently studied in Ref.~\cite{Becattini:2017ljh} for the real scalar field where it was shown that the fluid has a minimal proper temperature $T_U = a/2\pi$ where $a$ is the magnitude of the four-acceleration  vector ($a^\mu a_\mu = -a^2$). In this work we present a preliminary assessment of the case with spin by using the ansatz for the Wigner function presented in Ref.~\cite{Becattini:2013fla} and show that corresponding configurations are also solutions of the formalism developed in Ref.~\cite{Florkowski:2017ruc}.  

We start our presentation with a discussion of the unpolarized case in Sec.~\ref{sec:nospin}. In Sec.~\ref{sec:thermo} the local equilibrium distribution functions including spin polarization for the accelerating case are constructed (phase-space dependent spin density matrices). This, in Sec.~\ref{sec:spin}, allows us to introduce and solve the hydrodynamic equations with spin. In Sec.~\ref{sec:spinten} we give the appropriate forms of the spin tensor and the spin polarization vector, while in Sec.~\ref{sec:bound} we discuss the overall consistency of our approach which leads us to the constraint relating acceleration with the temperature. Finally, we summarize in Sec.~\ref{sec:summary}.

\begin{figure}[t]
\begin{center}
\includegraphics[angle=0,width=0.6\textwidth]{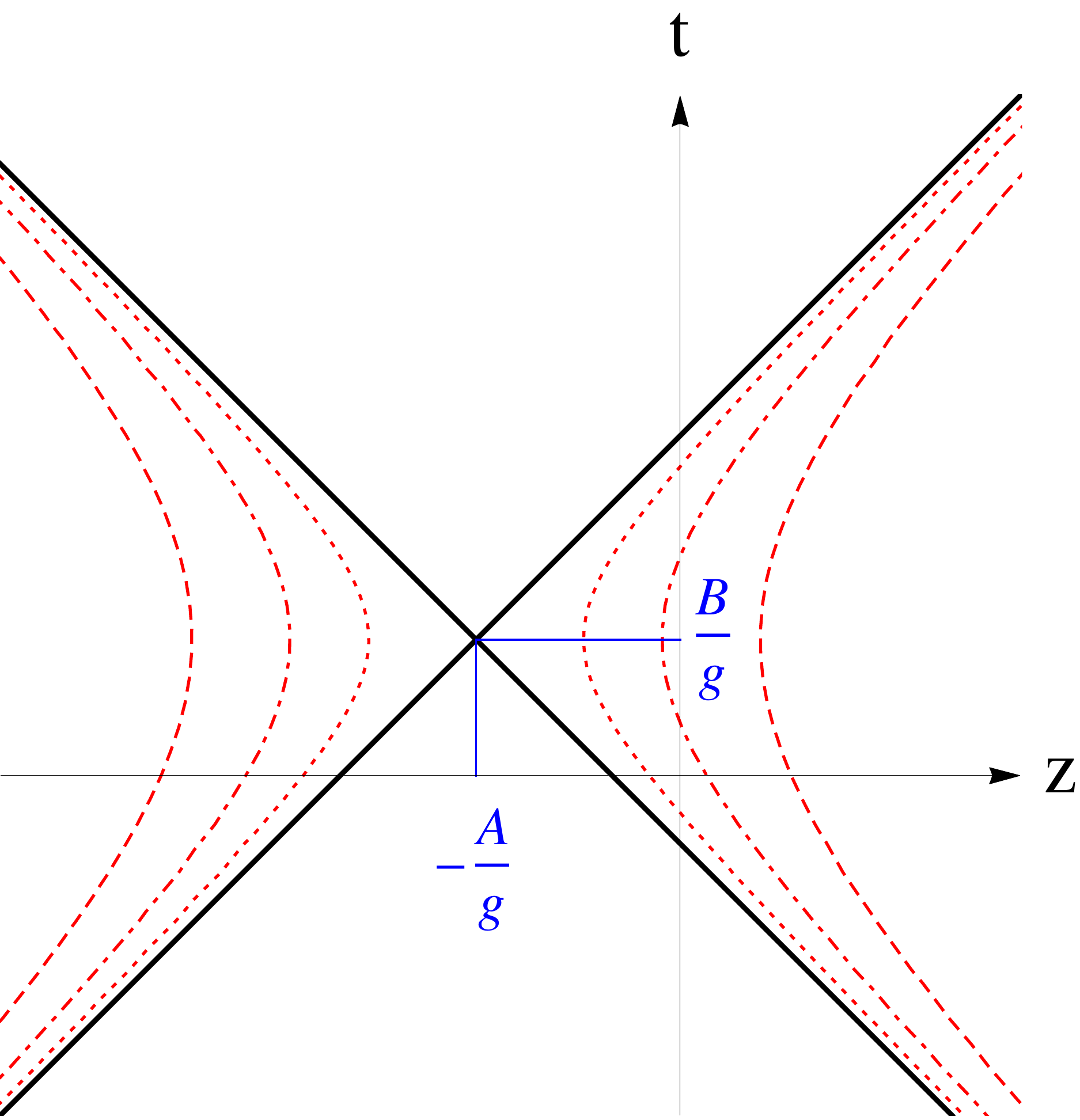} 
\caption{(Color online)  Space-time diagram showing trajectories (red dashed lines) of the fluid elements moving with a constant acceleration along the stream lines for the flow defined by \rf{u}. }
\label{fig:space-time}
\end{center}
\end{figure}
%

\section{The case without spin polarization}
\label{sec:nospin}

\subsection{Hydrodynamic flow}

In this work we consider a special case of one-dimensional hydrodynamic expansion with the fluid four-velocity $u^\mu$ introduced in Ref.~\cite{Becattini:2017ljh},
\bel{u}
u^\mu =  \gamma (1, 0,0, v) = \f{1}{\tau} \left(A+g z, 0, 0, -(B-g t) \right),
\eel
with
\bel{tau}
\tau = \sqrt{(A+g z)^2 - (B- g t)^2},
\eel
which guarantees the normalization $u^\mu u_\mu = 1$. With positive $A, B$, and $g$, we demand that $u^0 = \gamma > 0$ and $\tau > 0$. These two conditions specify the allowed region of spacetime where our fluid can expand: $z > -A/g$, $t > -z - (A-B)/g$, and $t < z + (A+B)/g$.  It corresponds to a quarter of the space-time diagram, placed to the right of the spacetime point with the coordinates $(B/g,0,0,-A/g)$, see Fig.~\ref{fig:space-time}.

It is easy to check that for any function of the variable $\tau$, let us say $F=F(\tau)$, its derivative along the fluid stream lines vanishes, namely
\bel{dF}
u^\mu \p_\mu F(\tau) = 0.
\eel
This means, in particular, that the lines of constant $\tau$ describe the stream lines of the fluid. Using \rf{dF} we further find the fluid four-acceleration
\bel{amu}
a^\nu = u^\mu \p_\mu u^\nu = \f{g}{\tau} \left(u^3, 0,0, u^0 \right),
\eel
which gives
\bel{a2}
a^\mu a_\mu = - \f{g^2}{\tau^2} \equiv -a^2, \quad a^\mu u_\mu = 0.
\eel
We thus conclude that the four-accelaration is constant along the stream lines.
It is also straightforward to see that the flow \rfn{u} is divergence free,
\bel{divu}
\p_\mu u^\mu = 0.
\eel
This result will be used below in the discussion of the conservation laws. 

\subsection{Perfect-fluid equations}

Before we turn to a discussion of spin polarization, let us demonstrate that the hydrodynamic flow of the form \rfn{u} is consistent with the evolution of the perfect fluid, provided its thermodynamic variables depend appropriately on the coordinates $t$ and $z$.
The energy momentum tensor of the perfect fluid has the form
\bel{Tmunu}
T^{\mu\nu} = (\varepsilon+P) u^\mu u^\nu - P g^{\mu\nu},
\eel
where $\ed$ and $P$ are the energy density and pressure. In this section we assume that both $\ed$ and $P$ are functions of the local temperature $T$ and chemical potential $\mu$, with the relations $\ed=\ed(T,\mu)$ and $P=P(T,\mu)$ specified by the fluid equation of state. The conservation law for the energy and momentum, $\p_\mu T^{\mu\nu} =0$, gives
\bel{dTmunu}
u^\nu u^\mu \p_\mu (\varepsilon+P) + u^\nu (\varepsilon+P) \p_\mu u^\mu + (\varepsilon+P) a^\nu = \p^\nu P.
\eel
In this section we  assume that $\ed$ and $P$ depend only on the variable $\tau$, namely
\bel{taudep}
\varepsilon = \varepsilon(T(\tau),\mu(\tau)), \quad P = P(T(\tau),\mu(\tau)).
\eel
In this case the first and second term in \rf{dTmunu} vanish, and we obtain 
\bel{dTmunu1}
(\varepsilon+P) a^\nu = \p^\nu P.
\eel

Using the standard thermodynamic relations
\bel{t1}
\ed+P = T s + \mu n
\eel
and
\bel{t2}
s = \left.{\f{\p P}{\p T}}\right\vert_{\mu}, \quad 
n = \left.{\f{\p P}{\p \mu}}\right\vert_{T},
\eel
where $s$ and $n$ are entropy and charge densities, respectively, the hydrodynamic equations \rfn{dTmunu1} can be rewritten as 
\bel{dTmunu2}
s \left(T a^\nu - \partial^\nu T\right) + n\left(\mu a^\nu - \partial^\nu \mu \right) = 0.
\eel
One can check that equations of the form 
\bel{X}
X a^\nu - \partial^\nu X = 0
\eel
are fulfilled if $X$ is of the form
\bel{Xsol}
X(\tau) = X_0 \f{\tau_0}{\tau}.
\eel
Thus, \rf{dTmunu2} has the solution of the form
\bel{Tmunusol}
T(\tau) = T_0 \f{\tau_0}{\tau}, \quad \mu(\tau) = \mu_0 \f{\tau_0}{\tau}.
\eel
We note that the conservation laws for entropy, $\p_\mu (s u^\mu) = 0$, as well as for the charge, $\p_\mu (n u^\mu)~=~0$, are also fulfilled due to Eqs.~\rfn{dF} and \rfn{divu}.
\\

By using Eqs.~(\ref{u}) and (\ref{Tmunusol}) we can show that the four-temperature field
$$
  \beta^\mu = \frac{u^\mu}{T} 
$$
is a Killing vector field, that is fulfilling the equation:
\bel{killing}
  \partial_\mu \beta_\nu + \partial_\nu \beta_\mu = 0,
\eel 
which shows that the only solution of ideal hydrodynamics with the given flow field (\ref{u}) 
and equilibrium thermodynamic relations is the {\em global} thermodynamic equilibrium one, 
with non-vanishing acceleration, as discussed in Ref.~\cite{Becattini:2017ljh}. As it is 
known, the general solution of the Killing equation in flat space-time reads:
\bel{beta}
\beta_\mu = b_\mu +\varpi_{\mu \nu} x^\nu,
\eel
where $b_\mu$ and $\varpi_{\mu \nu} = - \varpi_{\nu\mu}$ are constants. In our case, with 
the parametrization \rfn{u}, one has the following map:
\bel{b0b3om03}
b_0 = \f{A}{T_0 \tau_0}, \quad b_3 = \f{B}{T_0 \tau_0}, \quad 
\varpi_{03} = -\varpi_{30} = \f{g}{T_0 \tau_0}.
\eel
%

\section{Thermodynamic treatment of spin degrees of freedom}
\label{sec:thermo}

\subsection{Phase-space spin density matrices}

Recently, a new framework of relativistic hydrodynamics for particles with spin $\onehalf$ has been introduced in Refs.~\cite{Florkowski:2017ruc,Florkowski:2017dyn}. In this approach, the spin degrees of freedom are incorporated by using the phase-space density matrices for particles and antiparticles 
\cite{Becattini:2013fla}
\bel{fplusrsxp}
\fplusrsxp = \f{1}{2m} \ubarrp X^+ \usp,  \quad
\fminusrsxp = - \f{1}{2m} \vbarsp X^- \vrp.
\eel
Here $r,s = 1,2$ are spin indices, $u_r$ and $v_s$ are bispinors, and $X^{\pm}$ are the four by four matrices
\bel{XpmM}
X^{\pm} =  \exp\left[(\pm \mu - u_\mu \pmu ) /T \right] M^\pm,
\eel
where
\bel{Mpm}
M^\pm = \exp\left[ \pm \f{1}{2} \omnL(x)  \Sigma^{\mu \nu} \right].
\eel
Here $\omega_{\mu\nu}$ is the spin polarization tensor, and $\Sigma^{\mu \nu}$  is the spin operator expressed by the Dirac gamma matrices, $\Sigma^{\mu \nu}  = (i/4) [\gamma^\mu,\gamma^\nu]$.  The exponential functions appearing in \rfn{XpmM} and \rfn{Mpm} reflect the use of Boltzmann statistics for both particles and antiparticles. 

The explicit form of $M^\pm$ for arbitrary values of $\omega_{\mu\nu}$ has been recently worked out in \cite{Florkowski:2017dyn}
\beq
M^{\pm} &=&  \left[ \Re(\cosh Z) \pm \Re  \left( \frac{\sinh Z}{2 Z}\right) \omega_{\mu \nu}  \Sigma^{\mu \nu}\right]
\nonumber \\
&&\qquad\qquad+ i \gamma_5 \left[ \Im(\cosh Z)\pm \Im\left(\frac{\sinh Z}{2 Z}\right)\omega_{\mu \nu}   \Sigma^{\mu \nu} \right].
\label{MpmExp}
\eeq
Here $Z = \f{1}{2\sqrt{2}} 
\sqrt{\omega_{\mu\nu} \omega^{\mu\nu} + i \omega_{\mu\nu}  {\tilde \omega}^{\mu\nu}}$, where ${\tilde \omega}_{\mu\nu} = \f{1}{2} \varepsilon_{\mu\nu\alpha\beta} \omega^{\alpha\beta}$ is the dual tensor to $\omega_{\mu\nu}$.
If only the coefficients $\omega_{03}$ and $\omega_{30}$ are different from zero, one gets
\beq 
Z &=& \f{1}{2 \sqrt{2}} \sqrt{ \omnL \omnU}  =  \pm \f{i}{2} \omega_{03} .
\label{Z}
\eel
We stress that at this stage of our analysis we do not identify the spin polarization tensor $\omega_{\mu\nu}$ directly with the thermal vorticity $\varpi_{\mu\nu}$, see \rf{b0b3om03}, however, we assume that the only non-zero components of $\omega_{\mu\nu}$ and $\varpi_{\mu\nu}$ are  $\omega_{03} = - \omega_{30}$ and  $\varpi_{03} = - \varpi_{30}$.

In the calculation of $M^\pm$, the choice of the sign in \rf{Z} is irrelevant for $\cosh(Z)$ and $\sinh(Z)/Z$, thus in the following we select the upper sign. Using the notation
\bel{Zagain}
Z = i \zeta = i \f{\Omega}{T}
\eel
one gets
\beq
M^{\pm} &=&  \cos\left(\zeta\right) \pm  \frac{\sin\left( \zeta \right)}{2 \zeta}\omega_{\mu \nu}  \Sigma^{\mu \nu} .
\label{Mapp31} 
\eeq
It is interesting to note that \rf{Mapp31} is an analytic continuation of the expressions used before in Refs.~\cite{Florkowski:2017ruc,Florkowski:2017dyn} --- with real and positive $Z = \zeta = \Omega/T$ being continued to a purely imaginary value $Z = i \zeta = i \Omega/T$. The case of complex $Z$ was excluded from the investigations done in Refs.~\cite{Florkowski:2017ruc,Florkowski:2017dyn} as it may lead to negative values of the energy density. In fact, this is the situation we encounter in this work unless certain conditions on the parameters describing the motion of the fluid are imposed. We come back to a more detailed discussion of this point below in Sec.~\ref{sec:bound}.

\subsection{Basic physical observables}

Since the only difference in the calculation of thermodynamic variables, as compared to the case studied in Refs.~\cite{Florkowski:2017ruc,Florkowski:2017dyn},  is the change $\Omega \to i \Omega$, we can immediately use the results obtained before.
For the charge current, the replacement $\Omega \to i \Omega$ in Eq.~(11) of  Ref.~\cite{Florkowski:2017ruc} gives 
\bel{jmu}
N^\mu = n \umU,
\eel
where the charge density $n$ equals
\bel{nden}
n = 4 \, \cos(\zeta) \sinh(\xi)\, \n0(T). 
\eel
Here $\n0(T) = \langle(u\cdot p)\rangle_0$ is the number density of spinless, neutral Boltzmann particles, obtained from thermal averaging defined as
\bel{avdef}
\langle \cdots \rangle_0 \equiv \int \f{d^3p}{(2\pi)^3 E_p}  (\cdots) \,  e^{- \beta \cdot p} ,
\eel
with $p^0 = E_p = \sqrt{m^2 + {\boldsymbol p}^2}$ being the particle energy. In the similar way, starting from Eq.~(14) of Ref.~\cite{Florkowski:2017ruc}, we find the form of the energy-momentum tensor,
\bel{Tmn}
\TmnU  = (\varepsilon + P ) \umU \unu - P \gmunu,
\eel
with the energy density and pressure given by the formulas
\bel{enden}
\varepsilon = 4 \, \cos(\zeta) \cosh(\xi) \, \e0(T) 
\eel
and
\bel{prs}
P = 4 \, \cos(\zeta) \cosh(\xi) \, \P0(T), 
\eel
where $\e0(T) = \langle(u\cdot p)^2\rangle_0$ and $\P0(T) = -(1/3) \langle \left[ p\cdot p - 
(u\cdot p)^2 \right] \rangle_0$. 

\medskip
For the entropy current we use Eq.~(17) of Ref.~\cite{Florkowski:2017ruc} and find
\bel{s2}
S^\mu = s u^\mu,
\eel
with the entropy density given by the expression
\bel{s}
s = u_\mu S^\mu = \f{\ed+P  - \mu \, n - \Omega w}{T} , 
\eel
where
\bel{w}
w = -4 \, \sin(\zeta) \cosh(\xi) \, \n0(T). 
\eel
We note that in the case of the entropy density $s$ the situation is slightly different from the case of the thermodynamic variables $n$, $\ed$, and $P$, where the main effect of the analytic continuation is to replace $\cosh(\zeta)$ simply by $\cos(\zeta)$. The analytic continuation $\Omega \to i \Omega$ changes the sign of the original product of $\Omega \, w$ derived in Ref.~\cite{Florkowski:2017ruc}, since $\Omega \sinh{\zeta} \to - \Omega \sin{\zeta}$. Nevertheless, we can keep the same sign in front of the product $\Omega \,w$ in \rf{s}, as in Ref.~\cite{Florkowski:2017ruc}, if we introduce the minus sign in the definition of the new density~$w$, see \rf{w}. As the matter of fact, such a change of sign is consistent from the thermodynamic point of view, since with \rf{w} we can use, in addition to \rf{s},  the following thermodynamic relations
\bel{dP}
s = \left.{\f{\p P}{\p T}}\right\vert_{\mu,\Omega}, \quad 
n = \left.{\f{\p P}{\p \mu}}\right\vert_{T,\Omega}, \quad 
w = \left.{\f{\p P}{\p \Omega}}\right\vert_{T,\mu}. 
\eel
Equations \rfn{s} and \rfn{dP} become natural extensions of Eqs.~\rfn{t1} and \rfn{t2}.

\section{Hydrodynamic equations with spin}
\label{sec:spin}

In order to include spin polarization into hydrodynamic picture we use again \rf{dTmunu1}, which followed directly from the definition of the energy-momentum tensor and assumed geometry of expansion, and employ extended thermodynamic relation
\bel{t2n}
\ed+P = T s + \mu n + \Omega w.
\eel
This leads to the equation
\bel{dTmunu3}
s \left(T a^\nu - \partial^\nu T\right) + n\left(\mu a^\nu - \partial^\nu \mu \right)
 + w \left(\Omega a^\nu - \partial^\nu \Omega \right) = 0.
\eel
Hence, in addition to \rfn{dTmunu2} we should have
\bel{Omega}
\Omega = \Omega_0 \f{\tau_0}{\tau}.
\eel
The results for $T$ and $\Omega$ imply that $\omega_{03}$ is a constant. Indeed, using Eqs.~\rfn{Z} and \rfn{Zagain} one finds
\bel{omega03}
\omega_{03} = \f{2 \Omega}{T} =  \f{2 \Omega_0}{T_0} = \hbox{const}.
\eel
This result allows us to find the solutions with the spin polarization tensor equal to the thermal vorticity. In this case we demand that
\bel{cond1}
\varpi_{03} = \f{g}{T_0 \tau_0} = \omega_{03} =  \f{2 \Omega_0}{T_0},
\eel
which leads to
\bel{cond2}
\f{g}{\tau_0} = 2 \Omega_0.
\eel

\section{Spin tensor and spin polarization vector}
\label{sec:spinten}

For completeness, in this section we give the expressions for the spin tensor and the spin polarization vector.
The analytic continuation of Eq.~(17) in Ref.~\cite{Florkowski:2017ruc} together with the definition \rfn{w} allow us to write
the formula for the spin tensor as
\bel{spint}
S^{\lambda, \mu\nu} = -\f{w u^\lambda}{2 \zeta} \omega^{\mu\nu}.
\eel
In the case of symmetric energy-momentum tensor, which we consider in this work, the spin tensor is conserved,  $\p_\lambda S^{\lambda, \mu\nu} = 0$. This can be explicitly verified for the accelerating solution: in this case $\p_\lambda (w u^\lambda) = 0$, while $\zeta$ and $\omega^{\mu\nu}$ are constants (with $\omega_{03} = -\omega_{30}$ being the only non zero components of $\omega_{\mu\nu}$).

Similarly, using  expressions derived in  Ref.~\cite{Florkowski:2017dyn} we find the average polarization vector ${\boldsymbol {\cal P}}$ of particles with the mass $m$ and three-momentum ${\boldsymbol p}$,
\bel{calP}
{\boldsymbol {\cal P}} = \f{1}{2} \tan\left( \f{g}{2 T_0 \tau_0} \right) \f{{\boldsymbol p} \times {\hat {\boldsymbol z}} }{m}.
\eel
Here ${\hat {\boldsymbol z}}$ is a unit vector pointing in the direction of the $z$-axis.

\section{Lower bound on the temperature}
\label{sec:bound}

In Refs.~\cite{Florkowski:2017ruc,Florkowski:2017dyn} the two assumptions restricting the form of the spin polarization tensor $\omega_{\mu\nu}$ were made: $\omega_{\mu\nu} \omega^{\mu\nu}  > 0$ and  $\omega_{\mu\nu}  {\tilde \omega}^{\mu\nu} = 0$. They were motivated by the positivity condition for the energy density. In this work, the latter condition is fulfilled, however the former is not, since for the accelerating solution studied here we have  $\omega_{\mu\nu} \omega^{\mu\nu} < 0$. In fact, the expression \rfn{enden}, obtained for the energy density, contains the oscillating function $\cos(\zeta)$ that may have negative values. Therefore, although the framework constructed above is consistent from the thermodynamic and hydrodynamic points of view, we have to impose further restrictions on the parameters used in our scheme, which guarantee the positivity of the energy density.

The simplest way to assure that $\varepsilon > 0$ is by demanding that  $\zeta \leq \f{\pi}{2}$. For the accelerating solution, the quantity $\zeta$ is a constant, $\zeta = \f{\Omega_0}{T_0} = \f{g}{2 T_0 \tau_0} $, thus we find
\bel{constraint1}
\f{g}{T_0 \tau_0}  \leq \pi
\eel
or
\bel{constraint2}
\f{\sqrt{-a^\mu a_\mu}}{\pi} = \f{a}{\pi}  \leq T.
\eel

For $T = a/\pi$, we also get the pressure and charge density vanishing,
according to Eqs.~(\ref{nden}) and (\ref{prs}). This may suggest that $T=a/\pi$ corresponds
to the Minkowski vacuum for the spin-$\onehalf$ field just like $T=a/2\pi$ is for the
scalar field and that this is a limiting temperature for the fluid with spin. 
However, the spin tensor, which is proportional to $w$ (see eq.~(\ref{w})) 
does not vanish for $\zeta = \pi$, in fact, it is maximally negative. This indicates
that the condition $T=a/\pi$ cannot be interpreted as being equivalent to the true 
vacuum of the field, like in the Unruh case \cite{Becattini:2017ljh,Unruh:1976db}. This, together
with the fact that the temperature value differs by a factor of two, suggests that the 
occurrence of these limiting values are most likely related to the approximate character 
of the equilibrium distribution~\rfn{XpmM}. Nevertheless, we believe that they are
symptoms of an Unruh-like behaviour of the Dirac field at finite temperature and
acceleration which would most likely be in full agreement with that of the scalar field
in an exact calculation.

\section{Summary}
\label{sec:summary}

In this work we have presented solutions of a recently introduced framework of hydrodynamics with spin $\onehalf$, which describe the motion with constant acceleration along the stream lines. We have showed that the expansion of the fluid agrees with an example of the global thermodynamic equilibrium with non-zero polarization, introduced before in Ref.~\cite{Becattini:2017ljh}.

\bigskip
{\bf Acknowledgments:} 
W.F. and E.S. would like to thank Francesco Becattini for his kind hospitality in Florence where most of this work was done. 
W.F. was supported in part by the Polish National Science Center Grant No. 2016/23/B/ST2/00717. E.S. was supported by BMBF Verbundprojekt 05P2015 - Alice at High Rate. E.S. acknowledges partial support by the Deutsche Forschungsgemeinschaft (DFG) through the grant CRC-TR 211 ``Strong-interaction matter under extreme conditions". This work has been performed in the framework of COST Action CA15213
``Theory of hot matter and relativistic heavy-ion collisions" (THOR).

\end{document}